\documentclass[epj,twocolumn]{webofc}
\usepackage[varg]{txfonts}   
\woctitle{MPGD2015}
\begin{document}
\title{Micromegas for dark matter searches: CAST/IAXO \& TREX-DM experiments }

\author{J.~G.~Garza\inst{1}\fnsep\thanks{\email{jgraciag@unizar.es}} \and
        S.~Aune\inst{2} \and
        F.~Aznar\inst{1} \fnsep\thanks{Present address: Centro Universitario de la Defensa, Universidad de Zaragoza, Zaragoza, Spain} \and
  	J.~F.~Castel\inst{1}  \and
	S.~Cebri\'an\inst{1}  \and
	T.~Dafni\inst{1}  \and
	E.~Ferrer-Ribas\inst{2} \and
	J.~Gal\'an\inst{1} \and
	J.~A.~Garc\'ia\inst{1}  \and
	I.~Giomataris\inst{2} \and
	F.J.~Iguaz\inst{1}  \and
	I.G.~Irastorza\inst{1}  \and
	G.~Luz\'on\inst{1}  \and
	H.~Mirallas\inst{1}  \and
	T.~Papaevangelou\inst{2} \and
	A.~Peir\'o\inst{1}  \and
	A.~Tom\'as\inst{3} \and
	T.~Vafeiadis\inst{4} 
}

\institute{Grupo de F\'isica Nuclear y Astropart\'iculas, Universidad of Zaragoza, Zaragoza, Spain.
\and
Centre d'\'Etudes de Saclay, CEA, Gif-sur-Yvette, France.         
\and
Brackett Laboratory, Imperial College, London, UK.
\and
CERN, European Organization for Particle Physics and Nuclear Research, Geneva, Switzerland.
          }

\abstract{%
  The most compelling candidates for Dark Matter to day are WIMPs and axions. The applicability of gasesous Time Projection Chambers (TPCs) with Micromesh Gas Structures (Micromegas) to the search of these particles is explored within this work. Both particles would produce an extremely low rate at very low energies in particle detectors.  Micromegas detectors can provide both low background rates and low energy threshold, due to the high granularity, radiopurity and uniformity of the readout. Small (few cm wide) Micromegas detectors are used to image the axion-induced x-ray signal expected in the CERN Axion Solar Telescope (CAST) experiment. We show the background levels obtained in CAST and the prospects to further reduce them to the values required by the Internation Axion Observatory (IAXO). We also present TREX-DM, a scaled-up version of the Micromegas used in axion research, but this time dedicated to the low-mass WIMP detection. TREX-DM is a high-pressure Micromegas-based TPC designed to host a few hundreds of grams of light nuclei (argon or neon) with energy thresholds potentially at the level of 100~eV. The detector is described in detail, as well as the results of the commissioning and characterization phase on surface. Besides, the background model of TREX-DM is presented, along with the anticipated sensitivity of this search, which could go beyond current experimental limits.
}
\maketitle
%
\section{Micromegas in the search of Dark Matter: axions and WIMPs}\label{sec:intro}

The development of Micromegas-based gaseous Time Projection Chambers (TPCs) for the search of rare events has been the goal of the ERC-funded T-REX project~\cite{Irastorza2011_EAS}. In this document we present the application of this concept to the search of two of the most compelling dark matter candidates: axions and WIMPs, which arise in independent extensions of the Standard Model.

Although the detection of axions and WIMPs is conceptually very different, both searches share the extreme low rates and low energy of the expected signal events. The experimental challenge is therefore to achieve background rates and energy thresholds as low as possible. 

The mainstream of WIMP experiments has impressively increased their sensitivity thanks to the use of very large target masses and the capability to distinguish between electron and neutron-induced recoils. However, the efficiency of this discrimination technique is reduced below a relatively large energy, which implies a lost of sensitivity to low-mass WIMPs. Due to the intrinsic charge amplification happening in gaseous detectors, Micromegas are a promising technology to achieve very low energy thresholds. Although electron/nuclear discrimination in atmospheric gaseous TPCs is modest, high sensitivity to light WIMPs could be reached. TREX-DM is a prototype that pretends to test this detection concept. The apparatus, commissioning and first results are presented in section~\ref{sec:trexdm}.

On the other hand, if axions exist, they would be copiously produced in the Sun's core with $\sim$keV energies and they could be converted back to detectable x-rays in a laboratory magnetic field aligned with the Sun. The expected signal is  an excess of x-rays at the exit of the magnet over the background measured at non-alignment periods. Thus, ultra-low background x-ray detectors are a technological pillar of axion helioscopes. Here, we present the small Micromegas TPCs used in the CERN Axion Solar Telescope (CAST) since 2003, where they pioneered the application of these readouts to rare event searches. The background reduction techniques and the rates achieved are reviewed in section~\ref{sec:cast}, as well as the prospects for using this detection system for the future International Axion Observatory (IAXO)~\cite{Irastorza:1567109}, the next generation axion helioscope, now under proposal.

\section{Low background x-ray detection in CAST}\label{sec:cast}

One of the experimental challenges of axion helioscopes is the detection of a very low x-ray flux in the keV energy range. For this, the use of low background x-ray detection techniques is needed, potentially coupled to x-ray optics to focalize the parallel beam of photons into a small spot, in order to further increase the signal-to-noise ratio.

  The concept of detecting low energy x-rays with Micromegas detectors has been extensively developed and used in CAST since an early stage of the experiment~\cite{Abbon:2007ug,Aune:2013pna,Aune:2013nza}. The detector is a small TPC with a Micromegas readout at the anode, and whose cathode faces the magnet bore from where x-rays enter the detector. The conversion volume of the chamber is set to efficiently absorb the signal photons, while minimizing background, and typically has 3~cm height and is filled with 1.4 bar argon in addition to a small quantity of quencher. The x-rays coming from the magnet enter the conversion volume via a gas-tight window made of 5~$\mu$m aluminized mylar foil. This foil is also the cathode of the TPC, and is supported by a metallic strong-back, in order to withstand the pressure difference with respect to the magnet's vacuum system.

The ionization produced by the interactions in the conversion volume drifts and is projected onto the Micromegas readout plane, where signal amplification takes place. The Micromegas active area is 36~cm$^2$ that comfortably covers the 14.55~cm$^2$ projection of the magnet's bore, which fixes the analysis fiducial area. The readout plane is finely pixelized with a typical pitch of $\sim$0.5~mm. The pixels are read out in X and Y columns using state-of-the-art TPC data acquisition (DAQ) electronics based on the AFTER chip~\cite{Baron:2008zza}. This provides relative time-of-flight information of the ionization charge arriving to the readout plane and hence 3D information of the primary ionization cloud.

The sources of background for the Micromegas detectors are:  $\gamma$-rays, cosmic rays, the presence of radon around the detector, intrinsic radioactivity of the detector components or shielding, neutrons and cosmogenic activation of the detector materials. All these radiations can produce secondary or fluorescence emissions that can reach the detector active volume. The strategies to reduce these backgrounds are: the use of high-Z material (lead, copper) passive shielding to block the pass of $\gamma$-rays; active shielding to tag muons, such as plastic scintillators; the continuous flux of vaporized LN$_2$ into the detector environment to avoid the presence of air-borne radon; neutron moderator and absorber; and the use of radiopure components.

Muon-induced events are an important part of the background level of the Micromegas detectors when operated at surface level. In order to suppress these events, an active detector can be installed over the Micromegas detectors to act as a muon veto. The strategy is to use the time difference between the signal in the muon veto and the delayed Micromegas trigger to tag muon-induced events. A plastic scintillator acting as a muon veto was installed over the detectors for the data taking of 2012, and it was subsequently upgraded in 2013 due to its promising results. 

\begin{figure}
\centering
\includegraphics[width=7cm,clip]{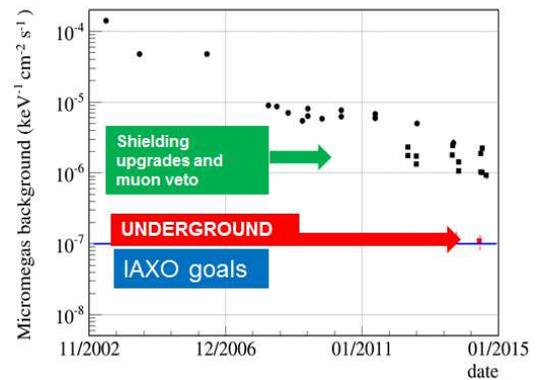}
\caption{Evolution of the background level in CAST, the limit obtained in the LSC and IAXO goals in the 2-7 keV energy range.}
\label{fig:historic}   
\end{figure}

The application of these techniques along with the reliability of the microbulk technology, the upgrade of the readout electronics, the high granularity of the readout -- which offers topological information of the event, a powerful tool for signal/background discrimination-- and the tunning of the rejection algorithms led to a background level reduction of a factor 10$^2$ in CAST-MM detectors over the last ten years. Figure~\ref{fig:historic} shows a compilation of background levels achieved in CAST in the keV energy range, at surface and underground test benches. The best level achieved at surface is already below 10$^{-6}$~keV$^{-1}$cm$^{-2}$s$^{-1}$~\cite{Aune:2013nza}. In particular, the best detector of CAST has demonstrated a background level as low as (0.8 $\pm$ 0.1)~keV$^{-1}$cm$^{-2}$s$^{-1}$ during the last data-taking campaigns. The background energy spectrum is characterized  by a fluorescence peak at 8~keV (from copper K$_{\alpha}$ emission), its escape peak at 5~keV and by the argon K$_{\alpha}$ line at 3~keV.

The operation of a replica of the detector in the Canfranc Underground Laboratory (LSC) sets a level as low as $\sim$10$^{-7}$~keV$^{-1}$cm$^{-2}$s$^{-1}$, almost at the required IAXO levels. According to our background model, based on in-situ measurements at CAST, tests underground and at surface as well as on Geant4 simulations, the contributions from $\gamma$-rays, radon and internal radioactivity have been reduced to negligible levels. The dominant contribution is due to cosmic muons and their secondary products, generated after their interaction in the setup components. On the other hand, the origin of the events limiting the LSC performance is uncertain. Some hypotheses are the $\beta$-decay of $^{39}$Ar present in the detection medium, neutrons or the cosmic activation of the materials. 

\begin{figure*}[Ht]
\centering
\includegraphics[width=13.5cm,clip]{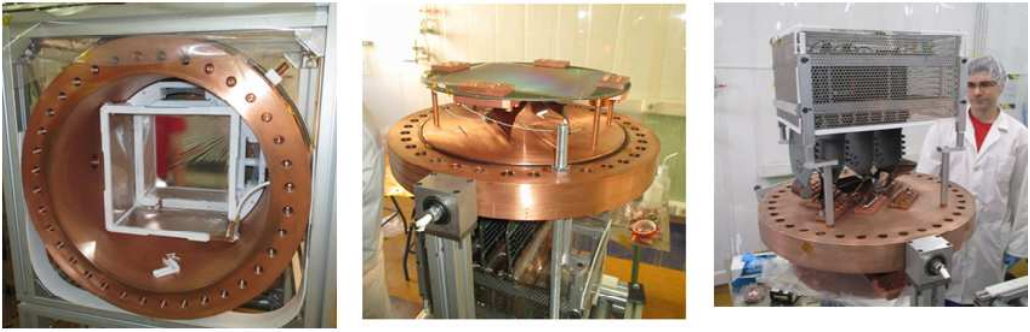}
\caption{Different views of the TREX-DM setup showing the field cage and cathode, Micromegas readout and electronics.}
\label{fig-setup}       
\end{figure*}

\subsection{Prospects for IAXO}\label{sec:prospects1}

The developments described in previous sections have a strong motivation as part of the preparatory activities for IAXO, the future next-generation axion helioscope now under proposal~\cite{Irastorza:1567109}. As part of this effort, we plan to build a prototype setup, IAXO-D0, with the aim of demonstrating the levels of background needed by IAXO.

This prototype is conceived as a generic test-bench for testing other technologies. The main improvement with respect to the CAST-MM is that a thicker and more compact shielding (~20--30~cm) and a 4$\pi$ enlarged muon veto system will be implemented. Pushing the lowest underground limit requires a replacement of the target gas by xenon or depleted argon from underground sources, or the installation of a neutron shielding. These activities are being developed in the context of the R\&D phase for the IAXO technical design report.

\section{TREX-DM prototype and characterization}\label{sec:trexdm}

In some respects, TREX-DM~\cite{trexdm_paper} is a scaled-up version of the CAST-MM, but with a 10$^{3}$ times larger active mass. Figure~\ref{fig-setup} shows different views of the apparatus. The vessel is comprised of a copper sleeve of 0.5~m diameter and 0.5~m length and two 6~cm thick flat end caps that can hold up to 12~bar. The inner volume is divided in two active volumes, separated by a central cathode made of 4~$\mu$m aluminized mylar foil. Around each active volume there is a 19~cm long and 25~cm wide square section field shaper that guarantees the homogeneity of the drift field, independently of the applied voltages in the cathode and mesh. The field shaper is composed of copper strips printed on single-layer of kapton, each strip separated by a 10~M$\Omega$ resistor. 

The Micromegas detectors are built using the bulk technology, with an amplification field of 128~$\mu$m. The active surface is 25.2$\times$25.2~cm$^{2}$, divided in square pads of 312~$\mu$m length with a pitch of 582~$\mu$m. The pads are alternatively interconnected in 432 strips per axis, which are rooted to four connector prints at the PCB sides. A flat cable links each detector footprint to the electronics by means of a commercial 300-pin connector. The mesh signal is extracted from the vessel by a coaxial low-noise cable and a feedthrough, and it is subsequently amplified and the spectrum recorded. The mesh also triggers the acquisition of the strip pulses, which come out from the vessel through flat cables and feed the AFTER-based front-end cards.

A $^{109}$Cd source is used to calibrate the detector at 22.1 (K$_{\alpha}$) and 24.9 keV (K$_{\beta}$). The Micromegas detectors are characterized in argon- and neon-based mixtures at high pressure with this source in order to find the optimum point of operation for a physics run.

A bulk Micromegas detectors  was installed at each end of the symmetric TPC, and were simultaneously  from 1.2 to 10~bar in steps of 1~bar in terms of electron transmission, detector gain, energy threshold and gain uniformity, in Ar + 2\%iC$_{4}$H$_{10}$. This is the first systematic characterization of bulk Micromegas detectors above atmospheric pressure to the authors' knowledge.  A gas flow of of 3--5~l/h is kept during the measurements.

The spectra are characterized by a peak at around 22~keV and by a fluorescence emission at around 6.4 and 8~keV from the iron and copper components. The spectral parameters are defined through an iterative multi-Gaussian fit corresponding to the K$_{\alpha}$ and K$_{\beta}$ emission lines of the source and their escape peaks.

\begin{figure}
\centering
\includegraphics[width=7.cm,clip]{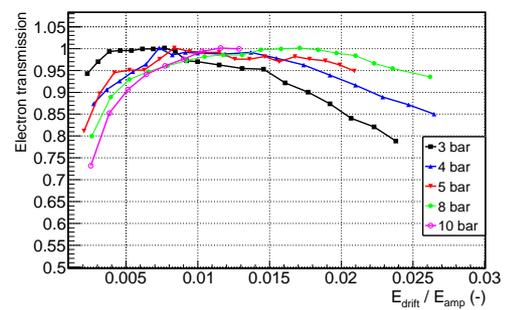}
\caption{Dependence of the electron transmission with the ratio of drift and amplification fields in Ar+2\%iC$_{4}$H$_{10}$.}
\label{fig-transp}   
\end{figure}

First, the drift voltage is varied for a fixed mesh voltage to obtain the electron transmission curve at each pressure (see figure~\ref{fig-transp}). The detectors show a plateau of maximum electron transmission for a wide range of ratios of drift and amplification fields at all pressures. The electron transmission drops at very low drift fields ($\sim$50~V/cm/bar) due to electron attachment and recombination of the primary electrons. For high drift fields, the mesh stops being transparent for primary electrons, and the energy resolution also degrades. 

\begin{figure}
\centering
\includegraphics[width=7.cm,clip]{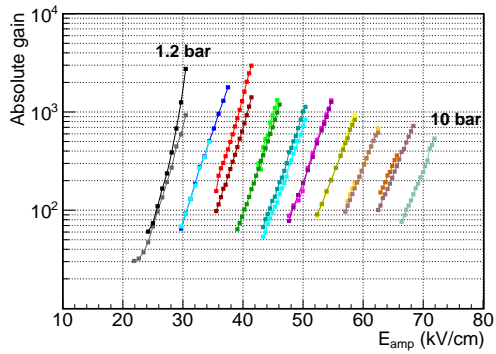}
\caption{ Dependence of the absolute gain with the amplification field between 1 and 10~bar, in steps of 1 bar. At each pressure the two Micromegas detectors of TREX-DM were characterized simultaneously.}
\label{fig-gain}   
\end{figure}

The absolute gain of the detector as a function of the amplification field is shown in figure~\ref{fig-gain} for all the pressure settings between 1 and 10 bar. The maximum gain decreases with pressure from $3\times10^{3}$ at 1.2 bar to  600 at 10~bar.

\begin{figure}
\centering
\includegraphics[width=7.cm,clip]{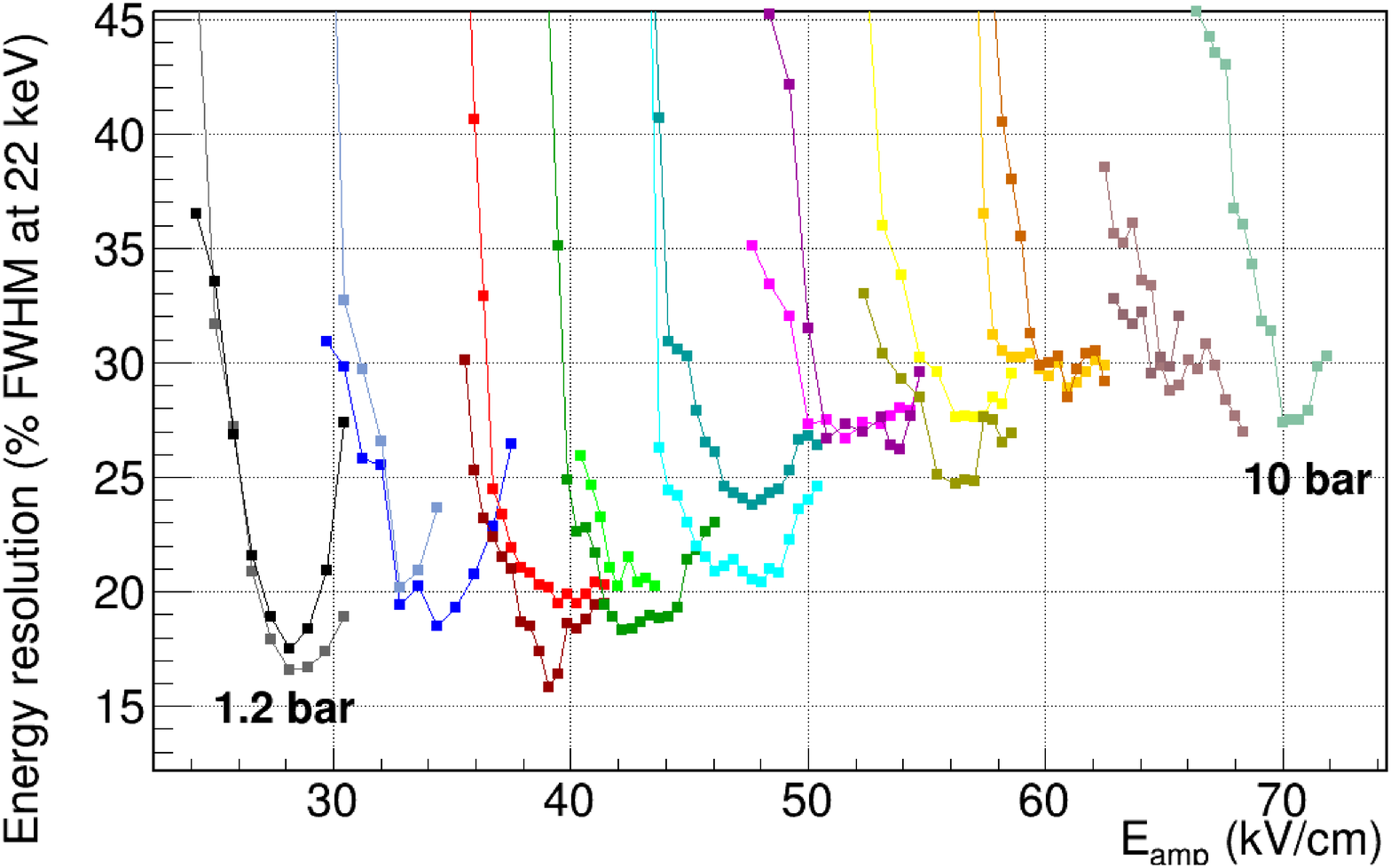}
\caption{Dependence of the energy resolution at 22~keV with the amplification field for the bulk Micromegas of TREX-DM in Ar+2\%iC$_{4}$H$_{10}$ between 1.2 and 10~bar. At each pressure the two Micromegas detectors of TREX-DM were characterized simultaneously.}
\label{fig-fwhm}   
\end{figure}

The dependence of the energy resolution with the amplification field for all the pressures settings is shown in figure~\ref{fig-fwhm}. At each pressure there is a range of amplification fields for which the energy resolution is optimized. At low gains, the energy resolution degrades because the signal becomes comparable with noise. At high fields, the resolution degrades due to increase in the gain fluctuations by the UV photons generated in the avalanche. As can be noticed, the best energy resolution degrades with the pressure, being around 16\% FWHM at 22.0~keV at 1.2~bar and above 25\%~FWHM at 10~bar; modest values for bulk detectors. The gain homogeneity over the detector surface has been also studied, showing gain fluctuations below 10\%.

\subsection{Prospects for light WIMP searches in TREX-DM}\label{sec:prospects2}

The role of the quencher fraction will be studied in near term. The modest gains achieved may be explained by the low quantity of quencher (2\%) in the gas.  A 5-10\% isobutane mixture will be tested in the near future, while neon-based mixtures will also be studied as they are expected to show hihger gains and consequently lower thresholds.

A first background model of the experiment in argon and neon-based mixtures in an underground environment has been developed in order to study the sensitivity of TREX-DM. Two target materials at 10~bar with an active mass of 0.3 and 0.16~kg have been considered, respectively for argon and neon.
The simulation process is described in detail in~\cite{trexdm_paper}. 

The expected background level for the argon-based gas is around $2 \times 10^2$~keV$^{-1}$kg$^{-1}$day$^{-1}$, dominated by the $^{39}$Ar isotope. If this contribution could be eliminated by using depleted argon from underground sources, the background level would be reduced to $\sim 1$~keV$^{-1}$kg$^{-1}$day$^{-1}$ in both argon and neon, limited by the copper vessel and connectors.

Assuming that the TREX-DM detector reaches an effective energy threshold of 0.4~keV and a conservative  background level of 10$^2$ keV$^{-1}$kg$^{-1}$day$^{-1}$, the experiment could be sensitive to a relevant fraction of the low-mass WIMP parameter space with a reasonable exposure. The experiment can set upper limits below to the low-mass ``region of interest'' invoked by some positive interpretations of some dark matter experiments.

\section*{Acknowledgements}
We thank our colleagues of CAST for many years of collaborative work in the experiment. We thank R.~de Oliveira and his team at CERN for the manufacturing of the microbulk readouts. We also thank the LSC staff for their help in the support of the Micromegas setup at the LSC. Authors would like to acknowledge the use of Servicio General de Apoyo a la Investigaci\'on-SAI, Universidad de Zaragoza. We acknowledge support from the European Commission under the European Research Council T-REX Starting Grant ref. ERC-2009-StG-240054 of the IDEAS program of the 7th EU Framework Program. We also acknowledge support from the Spanish Ministry of Economy and Competitiveness (MINECO) under contract ref.~FPA2013-41085, and under the CPAN project ref.~CSD2007-00042 from the Consolider-Ingenio 2010 program. Part of these grants are funded by the European Regional Development Fund (ERDF/FEDER).

\end{document}